\pdfoutput=1
\documentclass[journal]{IEEEtran}

\usepackage{graphicx}
\usepackage{psfrag}
\usepackage{epsfig}
\usepackage{ifthen}
\usepackage{law}
\usepackage{mystyle}
\usepackage{graphics}
\usepackage{times}
\usepackage{cite}

\usepackage{amsbsy} 
\usepackage{amsfonts} 
\usepackage{amsmath} 
\usepackage{amssymb} 
\usepackage{multicol}

\ifCLASSINFOpdf
\else
\fi

\begin{document}
%
\title{Spectrum Sensing for Cognitive Radio Using Kernel-Based Learning }

\author{\IEEEauthorblockN{Shujie~Hou and 
Robert C. Qiu, ~\IEEEmembership{Senior~Member,~IEEE}
\thanks{The authors are with the Department of Electrical and Computer Engineering, Center
for Manufacturing Research, Tennessee Technological University, Cookeville, TN 38505, USA. E-mail: shou42
@students.tntech.edu, rqiu@tntech.edu.}}}

\maketitle



%
\IEEEpeerreviewmaketitle

\begin{abstract}

Kernel method is a very powerful tool in machine learning. The trick of kernel has been effectively and extensively applied in many areas of machine learning, such as support vector machine (SVM) and kernel principal component analysis (kernel PCA). Kernel trick is to define a kernel function which relies on the inner-product of data in the feature space without knowing these feature space data. In this paper, the kernel trick will be employed to extend the algorithm of spectrum sensing  with leading eigenvector under the framework of PCA to a higher dimensional feature space. Namely, the leading eigenvector of the sample covariance matrix in the feature space is used for spectrum sensing without knowing the leading eigenvector explicitly. Spectrum sensing with leading eigenvector under the  framework of kernel PCA is proposed with the inner-product as a measure of similarity. A modified kernel GLRT algorithm based on matched subspace model will be the first time applied to spectrum sensing. The experimental results on simulated sinusoidal signal show that spectrum sensing with kernel PCA is about 4 dB better than PCA, besides, kernel GLRT is also better than GLRT. The proposed algorithms are also tested on the measured DTV signal. The simulation results show that kernel methods are 4 dB better than the corresponding linear methods. The leading eigenvector of the sample covariance matrix learned by kernel PCA is more stable than that learned by PCA for different segments of DTV signal.

\end{abstract}

\section{Introduction}

Spectrum sensing  is a cornerstone in cognitive radio ~\cite{haykin2005cognitive,mitola1999cognitive}, which detects the availability of radio frequency bands for possible use by secondary user  without interference to  primary user. Some traditional techniques proposed for spectrum sensing are energy detection, matched filter detection, cyclostationary feature detection,  covariance-based detection and feature based detection ~\cite{haykin2009spectrum, ma2009signal,cabric2004implementation,yucek2009survey,zeng2006maximum,zeng2008maximum,zeng2009spectrum,zeng2007covariance,zhang2011demonstration}. Spectrum sensing problem is nothing but a detection problem.

The secondary user receives the signal $y(t)$. Based on the received signal, there are two hypotheses: one is that the primary user is present ${\mathcal H}_1$, another one is the primary user is absent ${\mathcal H}_0$. In practice, spectrum sensing involves detecting whether the primary user is present or not from discrete samples of $y(t)$.
\begin{equation}
\label{eq1}
{ y(n)  =  }\left\{ \begin{array}{l}
 \begin{array}{*{20}c}
   {} & {}  \\
\end{array}{w(n)}\begin{array}{*{20}c}
   {\begin{array}{*{20}c}
   {} & {}  \\
\end{array}} & {} & {{\mathcal H}_0 }  \\
\end{array} \\ 
 {x(n) + w(n)}\begin{array}{*{20}c}
   {} & {} & {} & {{\mathcal H}_1}  \\
\end{array} \\ 
 \end{array} \right.
\end{equation}
in which $x(n)$ are  samples of the primary user's signal and $w(n)$ are samples of zero mean white Gaussian noise. In general, the algorithms of spectrum sensing aim at maximizing corresponding detection rate at a fixed false alarm rate with low computational complexity. The detection rate $P_d$ and false alarm rate $P_f$ are defined as 
\begin{equation}
\begin{array}{l}
 P_d  = prob({\rm {detect}}\:\: {\mathcal H}_1 |y(n) = x(n) + w(n)) \\ 
 P_f  = prob({\rm {detect}}\:\: {\mathcal H}_1 |y(n) = w(n)) \\ 
 \end{array}
\end{equation}
in which $prob$ represents probability.

Kernel methods ~\cite{schölkopf2002learning, weinberger2006unsupervised,lanckriet2004learning,cortes1995support} have been extensively and successfully applied in machine learning, especially in support vector machine (SVM) ~\cite{burges1998tutorial,smola2004tutorial}.  Kernel methods are counterparts of linear methods which implement in feature space. The data in original space can be mapped to different feature spaces with different kernel functions.  The diversity of feature spaces gives us more choice to gain  better performance's algorithm than only in the original space.

A kernel function which just relies on the inner-product of feature space data is defined as ~\cite{cristianini2000introduction}
\begin{equation}
\label{kernell}
{\mathop{\rm k}\nolimits} ({\bf x}_{i} ,{\bf x}_j ) = <\varphi ({\bf x}_{i} ), \varphi ({\bf x}_j )> 
\end{equation}
to implicitly map the original space data $\bf x$ into a higher dimensional feature space $\boldsymbol F$, where $\varphi$ is the mapping from original space to feature space. The dimension of $\varphi ({\bf x} )$ can be infinite, such as Gaussian kernel. Thus the direct operation on $\varphi ({\bf x} )$ may be computationally infeasible. However, with the use of the kernel function, the computation will only rely on the inner-product between the data points. Thus the extension of some algorithms to even an arbitrary dimensional feature space becomes possible.

$<{\bf x}_i ,{\bf x}_j >$ is the inner-product between ${\bf x}_i $ and ${\bf x}_j$. A function ${\mathop{\rm k}\nolimits}$ is a valid kernel if there exists a mapping $\varphi$ satisfying Eq. \eqref{kernell}. Mercer's condition ~\cite{cristianini2000introduction} gives us the condition  about what kind of functions are valid kernels. Kernel functions allow the linear method to generalize to a non-linear method without knowing $\varphi$ explicitly.  If the data in original space embodies nonlinear structure, kernel methods can usually obtain better performance than linear methods.

Spectrum sensing with leading eigenvector of the sample covariance matrix is proposed and hardware demonstrated in ~\cite{zhang2011demonstration} successfully under the framework of PCA. The leading eigenvector of non-white wide-sense stationary (WSS) signal has been proved stable ~\cite{zhang2011demonstration}.  In this paper, spectrum sensing with leading eigenvector of the sample covariance matrix of feature space data is proposed. The kernel trick is employed to implicitly map the original space data to a higher dimensional feature space.  In the feature space, the inner-product is taken as a measure of similarity between leading eigenvectors without knowing leading eigenvectors explicitly. That is to say spectrum sensing with leading eigenvector under the  framework of kernel PCA is proposed with the inner-product as a measure of similarity.

Several generalized likelihood ratio test (GLRT) ~\cite{lim2008glrt,font2010glrt} algorithms have been proposed for spectrum sensing.  Kernel GLRT ~\cite{kwon2006kernel} algorithm based on matched subspace model ~\cite{scharf1994matched} is proposed and applied to hyperspectral target detection problem, which assumes that the target and background lie in the known linear subspaces $[\bf T ]$ and $[\bf B]$. $\bf T$ and $\bf B$ are orthonormal matrices with the columns of each spanning the subspaces $[\bf T]$ and $[\bf B]$. $\bf T$ and $\bf B$ consist of  eigenvectors corresponding to nonzero eigenvalues of the sample covariance matrices of target and background, respectively. The identity  projection operator in the feature space is assumed to map $\varphi ({\bf x})$ onto the subspace consisting of the linear combinations of column vectors of $\bf T$ and $\bf B$ . 

 In this paper,  modified kernel GLRT algorithm based on matched subspace model  will be the first time employed for spectrum sensing without consideration of  background. On the other hand, the identity  projection operator in the feature space is assumed to map $\varphi ({\bf x})$ as $\varphi ({\bf x})$ in this paper.

The contribution of this paper is as follows: Detection algorithm with leading eigenvector will be generalized to feature spaces which are determined by the choice of kernel functions. Simply speaking, leading eigenvector detection based on kernel PCA is proposed for spectrum sensing. 
Different from PCA, the similarity of leading eigenvectors will be measured by inner-product instead of the maximum absolute value of cross-correlation.
A modified version of kernel GLRT  will be introduced to spectrum sensing which considers the perfect identity  projection operator in feature space without involving background signal. DTV signal ~\cite{Tawil} captured in Washington D.C. will be employed to test the proposed kernel PCA and kernel GLRT algorithms for spectrum sensing.

The organization of this paper is as follows.  In section~\ref{kpca},  spectrum sensing with leading eigenvector under the framework of PCA will be reviewed.  Detection with  leading eigenvector will be extended to feature space by use of  kernel. The proposed algorithm that spectrum sensing with  leading eigenvector under the framework of  kernel PCA will be introduced in section~\ref{kpca}.  GLRT and modified kernel GLRT algorithms for spectrum sensing based on matched subspace model will be introduced in section~\ref{kernelGLRT}. The experimental results on simulated sinusoidal signal and DTV signal are shown in section~\ref{experiments}. The corresponding kernel methods will be compared with linear methods. Finally, the paper is concluded in section ~\ref{conclusion}.

\section{Spectrum Sensing with PCA and Kernel PCA}
\label{kpca}

The $d-$dimensional received vector is ${\bf y} = (y(n),y(n + 1),...,y(n + d - 1))^T $, therefore,
\begin{equation}
\begin{array}{l}
 {\mathcal H}_0 :{\bf y} = {\bf w} \\ 
 {\mathcal H}_1 :{\bf y} = {\bf x} + {\bf w} \\ 
 \end{array}
\end{equation}
in which ${\bf x} = (x(n),x(n + 1),...,x(n + d - 1))^T $ and ${\bf w} = (w(n),w(n + 1),...,w(n + d - 1))^T $.
Assuming the samples of the primary user's signal is known priorly with length $L>d$, $x(n),x(n + 1),...,x(L-1)$.  The training set consists of 
\begin{equation}
\label{sample}
\begin{array}{ll}
{\bf x}_1= &(x(n),x(n + 1),...,x(n + d - 1))^T ,\\
{\bf x}_2= &(x(n+i),x(n +i+1),...,x(n + i+d - 1))^T, \\
&\cdots,\cdots, \\
{\bf   x}_M=& (x(n+(M-1)i),x(n + (M-1)i+1),\\
&...,x(n + (M-1)i+d - 1))^T,\\
\end{array}
\end{equation}
where $M$ is the number of vectors in the training set and $i$ is the sampling interval. $T$ represents transpose.

\subsection{Detection Algorithm with Leading Eigenvector under the Framework of PCA}

The leading eigenvector (eigenvector corresponding to the largest eigenvalue) of the sample covariance matrix of the training set can be obtained which is taken as the template of PCA method. Given $d$-dimensional column vectors ${\bf x}_1,{\bf x}_2, \cdots, {\bf   x}_M$ of the training set, the sample covariance matrix can be obtained by
 \begin{equation}
 \label{cov}
{\bf R}_x  = \frac{1}{M}\sum\limits_{i = 1}^M {{\bf x}_i } {\bf x}_{i }^T,
 \end{equation}
which assumes that the sample mean is zero,
\begin{equation}
{\bf u} = \frac{1}{M} \sum\limits_{i = 1}^M {{\bf x}_i }  = {\bf 0}.
\end{equation}
The leading eigenvector of ${\bf R}_x$ can be extracted by eigen-decomposition of  ${\bf R}_x$,
\begin{equation} 
{\bf R}_x  = {\bf V\Lambda V}^T 
\end{equation}
where ${\bf \Lambda } = diag(\lambda _1 ,\lambda _2 ,...,\lambda _d )$ is a diagonal matrix. $ \lambda _i,i=1,2,\cdots,d $ are eigenvalues of ${\bf R}_x$. $\bf V$ is an orthonormal matrix, the columns of which  ${\bf v}_1,{\bf v}_2,\cdots, {\bf v}_d$  are the eigenvectors corresponding to the eigenvalues $ \lambda _i,i=1,2,\cdots,d$.
For simplicity,  take ${\bf v}_1$ as the eigenvector corresponding to the largest eigenvalue. The leading eigenvector ${\bf v}_1$ is the template of PCA. 

For the received samples  $(y(n),y(n + 1),...,y(L-1))$, likewise,  vectors ${\bf y}_i,i=1,2,\cdots,M$ can be obtained by \eqref{sample}. (Indeed, the number of the training set is not necessarily equal to the number of the received vectors, here, for simplicity, we use the same $M$ to denote both of them.) The leading eigenvector ${\bf \tilde v}_1$ of the sample covariance matrix  ${\bf R}_y= \frac{1}{M}\sum\limits_{i = 1}^M {{\bf y}_i } {\bf y}_{i }^T$ is obtained. The presence of $x(n)$ in $y(n)$  is determined by  
\begin{equation}
\rho  = \mathop {\max }\limits_{l = 0,1,...,d} \left| {\sum\limits_{k = 1}^d {{\bf v}_1 [k]{\bf \tilde v}_1 [k + l]} } \right|> T_{pca},
\end{equation}
where $T_{pca}$ is the threshold value for PCA method, and $\rho$ is the similarity between ${\bf \tilde v}_1$  and template ${\bf v}_1$  which is measured by cross-correlation. $T_{pca}$ is assigned to arrive a desired false alarm rate.
The detection with leading eigenvector under the framework of PCA is simply called PCA detection.

\subsection{Detection Algorithm with Leading Eigenvector under the Framework of Kernel PCA}

A nonlinear version of PCA--kernel PCA ~\cite{schölkopf1998nonlinear}-- has been proposed based on the classical PCA approach. Kernel function is employed by kernel PCA to implicitly map the data into a higher dimensional feature space, in which PCA is assumed to work better than in the original space.
By introducing the kernel function, the mapping $\varphi$ need not be explicitly known which can obtain better performance without increasing much computational complexity.

The training set ${\bf x}_i,i=1,2,\cdots,M$ and received set ${\bf y}_i,i=1,2,\cdots,M$ in kernel PCA are obtained  the same way as with PCA framework.

The training set in the feature space are $\varphi ({\bf x}_1 ),\varphi ({\bf x}_2 ),...,\varphi ({\bf x}_M )$ which are assumed to have zero mean, e.g., $\frac{1}{M} \sum\limits_{i = 1}^M {\varphi ({\bf x}_i )}  = {\bf 0}$. Similarly, the sample covariance matrix of $\varphi ({\bf x}_i )$ is 
\begin{equation}
{\bf R}_{\varphi (x)}  = \frac{1}{M} \sum\limits_{i = 1}^M {\varphi ({\bf x}_i )} \varphi ({\bf x}_i )^T .
\end{equation}
 The leading eigenvector ${\bf v}_1^f$ of ${\bf R}_{\varphi (x)}$ corresponding to the largest eigenvalue $ {\lambda}_1 ^f$ satisfies
 \begin{equation}
 \label{linear}
\begin{array}{ll}
 {\bf R}_{\varphi (x)} {\bf v}_1^f  &= {\lambda}_1 ^f {\bf v}_1^f  \\ 
  &= \frac{1}{M}\sum\limits_{i = 1}^M {\varphi ({\bf x}_i )\varphi ({\bf x}_i )^T } {\bf v}_1^f  = {\lambda}_1 ^f {\bf v}_1^f  \\ 
  &= \frac{1}{M}\sum\limits_{i = 1}^M { < \varphi ({\bf x}_i ),{\bf v}_1^f  > \varphi ({\bf x}_i )}  = {\lambda}_1 ^f {\bf v}_1^f. \\
 \end{array}
 \end{equation}
The last equation in \eqref {linear} implies that the eigenvector ${\bf v}_1^f$ is the linear combination of the feature space data $\varphi ({\bf x}_1 ),\varphi ({\bf x}_2 ),...,\varphi ({\bf x}_M )$,
\begin{equation}
\label{vect}
{\bf v}_1^f  = \sum\limits_{i = 1}^M {\beta _i \varphi ({\bf x}_i )}.
\end{equation}

Substituting \eqref{vect} into \eqref{linear},
\begin{equation}
\label{k_vec}
\frac{1}{M}\sum\limits_{i = 1}^M {\varphi ({\bf x}_i )\varphi ({\bf x}_i )^T } \sum\limits_{j = 1}^M {\beta _i \varphi ({\bf x}_j )}  = {\lambda}_1 ^f \sum\limits_{j = 1}^M {\beta _j \varphi ({\bf x}_j )} 
\end{equation}
and left multiplying  $\varphi ({\bf x}_t )^T,t=1,2,\cdots,M $ to both sides of \eqref{k_vec}, yields
\begin{equation}
\begin{array}{l}
\label{kernel_mat}
\frac{1}{M}\sum\limits_{i = 1}^M { < \varphi ({\bf x}_t ),\varphi ({\bf x}_i ) > } \sum\limits_{j = 1}^M {\beta _j  < \varphi ({\bf x}_i ),\varphi ({\bf x}_j )}  > \\ 
= {\lambda}_1 ^f \sum\limits_{j = 1}^M {\beta _j  < \varphi ({\bf x}_t ),\varphi ({\bf x}_j )}  > .\\
\end{array}
\end{equation}
By introducing the kernel matrix ${\bf K} = ({\mathop{\rm k}\nolimits} ({\bf x}_i ,{\bf x}_j ))_{ij} =  (< \varphi ({\bf x}_i ),\varphi ({\bf x}_j ) >)_{ij} $ and vector 
${\boldsymbol \beta }_1 = (\beta _1 ,\beta _2 ,....,\beta _M )^T $, eq. \eqref{kernel_mat} becomes 
\begin{equation}
 {\bf K}^2 {\boldsymbol \beta }_1  = M{\lambda}_1 ^f {\bf K}{\boldsymbol \beta }_1   => {\bf K {\boldsymbol \beta }_1 } = M{\lambda}_1 ^f {\boldsymbol \beta }_1.
\end{equation}
It can be seen that ${\boldsymbol \beta }_1$ is the leading eigenvector of the kernel matrix $\bf K$. The kernel matrix $\bf K$ is positive semidefinite.

Thus, the coefficients ${\beta}_i$ in \eqref{vect} for  ${\bf v}_1^f$ can be obtained by eigen-decomposition of the kernel matrix $\bf K$ which has been proved in ~\cite{schölkopf1998nonlinear} before.
The normalization of  ${\bf v}_1^f$  can be derived by ~\cite{schölkopf1998nonlinear}
\begin{equation}
\label{norm}
\begin{array}{ll}
 1 &=  < {\bf v}_1^f ,{\bf v}_1^f  >  \\ 
  &=  < \sum\limits_{i = 1}^M {\beta _i \varphi ({\bf x}_i )} ,\sum\limits_{i = 1}^M {\beta _i \varphi ({\bf x}_i )}  >  \\ 
  &= \sum\limits_{i,j = 1}^M {\beta _i \beta _j }  < \varphi ({\bf x}_i ),\varphi ({\bf x}_j ) >  \\ 
  &= {\boldsymbol \beta }_1^T {\bf K}{\boldsymbol \beta }_1 \\ 
  &= {\boldsymbol \beta }_1^T {\mu}_1 {\boldsymbol \beta }_1 \\ 
  &= {\mu}_1  < {\boldsymbol \beta }_1,{\boldsymbol \beta }_1 >  \\ 
 \end{array}
\end{equation}
in which ${\mu}_1$ is the eigenvalue corresponding to the eigenvector ${\boldsymbol \beta }_1$ of $\bf K$.

In the traditional kernel PCA approach ~\cite{schölkopf1998nonlinear}, the  first principal component of a random point $\varphi ({\bf x})$ in the feature space can be extracted by
\begin{equation}
\begin{array}{ll}
 < \varphi ({\bf x}),{\bf v}_1^f  >  &= \sum\limits_{i = 1}^M {\beta _i  < \varphi ({\bf x}),\varphi ({\bf x}_i ) > }\\
 &= \sum\limits_{i = 1}^M {\beta _i } {\mathop{\rm k}\nolimits} ({\bf x},{\bf x}_i ) ,\\
 \end{array}
\end{equation}
without knowing ${\bf v}_1^f$ explicitly. 

However, instead of computing principal components in the feature space, the leading eigenvector ${\bf v}_1^f $ is needed as the template for the detection problem. Though ${\bf v}_1^f $ can be written as the linear combination of $\varphi ({\bf x}_1 ),\varphi ({\bf x}_2 ),...,\varphi ({\bf x}_M )$ in which the coefficients are entries of leading eigenvector of $\bf K$, because  $\varphi ({\bf x}_1 ),\varphi ({\bf x}_2 ),...,\varphi ({\bf x}_M )$  are not given, the leading eigenvector ${\bf v}_1^f $ is still not explicitly known. 

In this paper, a detection scheme based on the leading eigenvector of the sample covariance matrix in the feature space is proposed without knowing ${\bf v}_1^f $ explicitly. 

Given the received vectors ${\bf y}_i,i=1,2,\cdots,M$, likewise, the leading eigenvector ${\bf \tilde v}_1^f $ of  the sample covariance matrix ${\bf R}_{\varphi (y)}$ is the linear combination of the feature space data $\varphi ({\bf y}_1 ),\varphi ({\bf y}_2 ),...,\varphi ({\bf y}_M )$, e.g.,
\begin{equation}
{\bf \tilde v}_1^f  = \sum\limits_{i = 1}^M {\tilde \beta _i } \varphi ({\bf y}_i ).
\end{equation}
${{\tilde {\boldsymbol \beta }}_1}=(\tilde \beta _1 ,\tilde \beta _2 ,...,\tilde \beta _M )^T $ is the leading eigenvector of the kernel matrix 
\begin{equation}
{\bf \tilde K} = ({\mathop{\rm k}\nolimits} ({\bf y}_i ,{\bf y}_j ))_{ij}  =  (< \varphi ({\bf y}_i ),\varphi ({\bf y}_j ) >)_{ij} .
\end{equation}

As is well-known that inner-product is one kind of similarity measure. Here, the similarity between ${\bf v}_{^1 }^f$ and ${\bf \tilde v}_1^f $ is measured by inner-product.
\begin{equation}
\label{sim}
\begin{array}{lll}
 \rho  &=&  < {\bf v}_{^1 }^f ,{\bf \tilde v}_1^f  >  =  < \sum\limits_{i = 1}^M {\beta _i \varphi ({\bf x}_i )} ,\sum\limits_{j = 1}^M {\tilde \beta _i } \varphi ({\bf y}_i ) >  \\ 
 \\
  &=& \{ (\varphi ({\bf x}_1 ),\varphi ({\bf x}_2 ),...,\varphi ({\bf x}_M )){\boldsymbol \beta }_1 \} ^T \cdot\\
  &&\{ (\varphi ({\bf y}_1 ),\varphi ({\bf y}_2 ),...,\varphi ({\bf y}_M )){\tilde {\boldsymbol \beta }}_1 \}  \\ 
  \\
 & =& {\boldsymbol \beta }_{_1 }^T \left( \begin{array}{c}
 \varphi ({\bf x}_1 )^T  \\ 
 \varphi ({\bf x}_2 )^T  \\ 
  \vdots  \\ 
 \varphi ({\bf x}_M )^T  \\ 
 \end{array} \right)(\varphi ({\bf y}_1 ),\varphi ({\bf y}_2 ),...,\varphi ({\bf y}_M )){\tilde {\boldsymbol \beta }}_1  \\ 
 \\
  &=& {\boldsymbol \beta }_{_1 }^T \left( \begin{array}{c}
 {\mathop{\rm k}\nolimits} ({\bf x}_1 ,{\bf y}_1 ),{\mathop{\rm k}\nolimits} ({\bf x}_1 ,{\bf y}_2 ),...,{\mathop{\rm k}\nolimits} ({\bf x}_1 ,{\bf y}_M ) \\ 
{\mathop{\rm k}\nolimits} ({\bf x}_2 ,{\bf y}_1 ),{\mathop{\rm k}\nolimits} ({\bf x}_2 ,{\bf y}_2 ),...,{\mathop{\rm k}\nolimits} ({\bf x}_2 ,{\bf y}_M ) \\ 
 ...... \\ 
 {\mathop{\rm k}\nolimits} ({\bf x}_M ,{\bf y}_1 ),{\mathop{\rm k}\nolimits} ({\bf x}_M ,{\bf y}_2 ),...,{\mathop{\rm k}\nolimits} ({\bf x}_M ,{\bf y}_M ) \\ 
 \end{array} \right){\tilde {\boldsymbol \beta }}_1  \\ 
 \\
 & =& {\boldsymbol \beta }_{_1 }^T {\bf K}^t {\tilde {\boldsymbol \beta }}_1.  \\ 
 \end{array}
\end{equation}

${\bf K}^t $ is the kernel matrix between $\varphi ({\bf x}_i )$ and $\varphi ({\bf y}_j )$. A measure of similarity between ${\bf v}_{^1 }^f$ and ${\bf \tilde v}_1^f $ has been obtained without giving  ${\bf v}_{^1 }^f$ and ${\bf \tilde v}_1^f $ based on  \eqref{sim}.

\begin{figure}[!t]
\begin{center}
\scalebox{.40}{\includegraphics{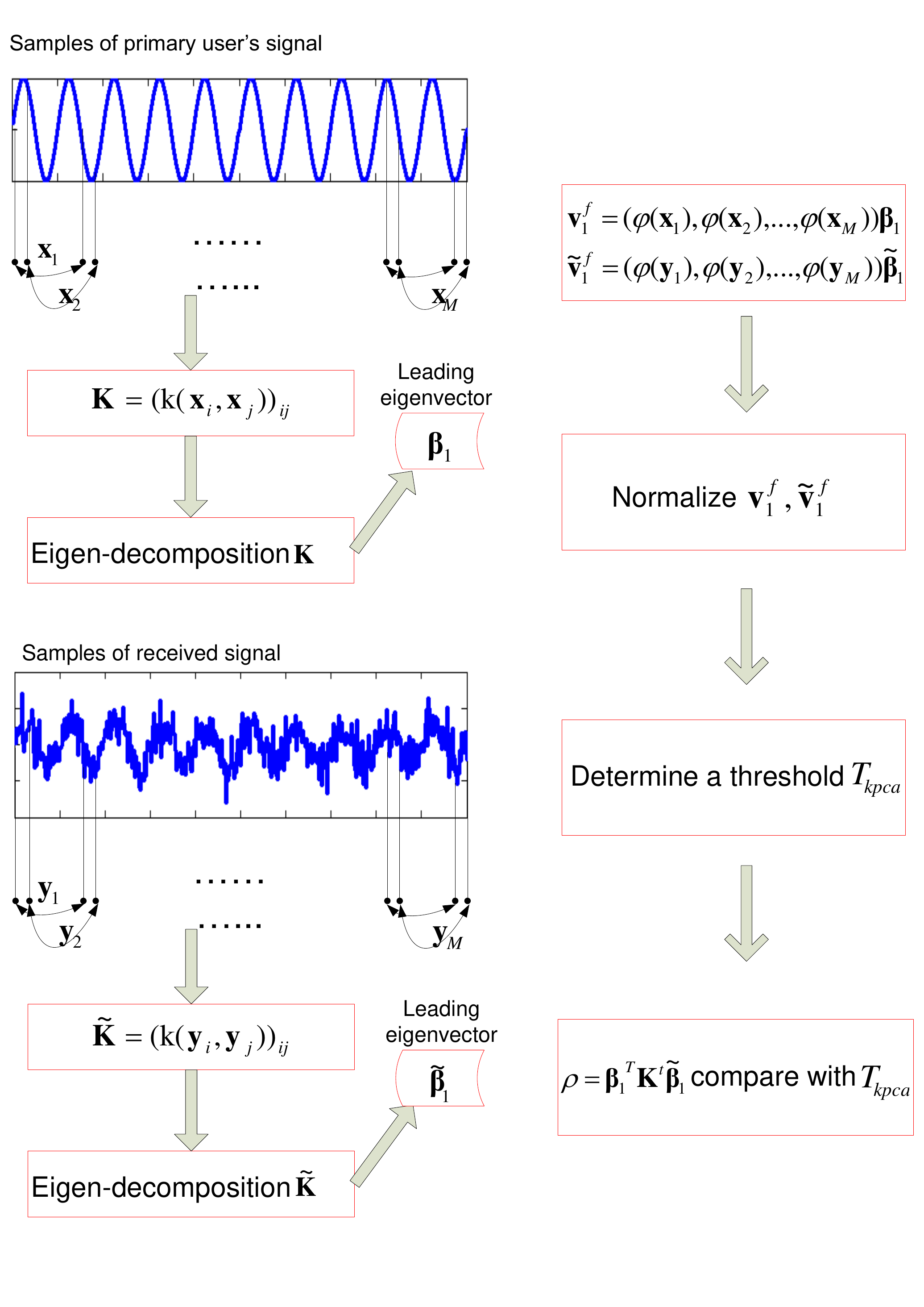}}
\end{center}
\vspace{-1ex}
\caption{The flow chart of the proposed kernel PCA algorithm for spectrum sensing }
\vspace{-2ex}
\label{fig:kpca}
\end{figure}

The proposed detection algorithm with  leading eigenvector under the framework of kernel PCA is summarized here as follows:
\begin{enumerate}
\item Choose a kernel function ${\mathop{\rm k}\nolimits}$. Given the training set of the primary user's signal  ${\bf x}_1,{\bf x}_2,\cdots,{\bf x}_M$, the kernel matrix is ${\bf K} = ({\mathop{\rm k}\nolimits} ({\bf x}_i ,{\bf x}_j ))_{ij}$. $\bf K$ is positive semidefinite. Eigen-decomposition of $\bf K$ to obtain the leading eigenvector ${\boldsymbol \beta }_{_1 }$.
\item The received vectors are ${\bf y}_1,{\bf y}_2,\cdots,{\bf y}_M$. Based on the chosen kernel function, the kernel matrix ${\bf \tilde K} = ({\mathop{\rm k}\nolimits} ({\bf y}_i ,{\bf y}_j ))_{ij} $ is obtained. The leading eigenvector ${\tilde {\boldsymbol \beta }}_1$ is also obtained by eigen-decomposition of ${\bf \tilde K}$.
\item The leading eigenvectors for ${\bf R}_{\varphi (x)} $ and ${\bf R}_{\varphi (y)}$ can be expressed as 
\begin{equation}
\begin{array}{l}
{\bf v}_{^1 }^f =(\varphi ({\bf x}_1 ),\varphi ({\bf x}_2 ),...,\varphi ({\bf x}_M )){\boldsymbol \beta }_1,\\
{\bf \tilde v}_1^f= (\varphi ({\bf y}_1 ),\varphi ({\bf y}_2 ),...,\varphi ({\bf y}_M )){\tilde {\boldsymbol \beta }}_1.
\end{array}
\end{equation}
\item Normalize ${\bf v}_{^1 }^f$ and ${\bf \tilde v}_1^f $ by \eqref{norm}.
\item The similarity between  ${\bf v}_{^1 }^f$ and ${\bf \tilde v}_1^f $ is 
\begin{equation}
\rho  = {\boldsymbol \beta }_{_1 }^T {\bf K}^t {\tilde {\boldsymbol \beta }}_1.
\end{equation}
\item Determine the presence or absence of primary signal  $x(n)$ in $y(n)$ by  evaluating $\rho>T_{kpca}$ or not.
\end{enumerate}
$T_{kpca}$ is the threshold value for kernel PCA algorithm.
The flow chart of the proposed kernel PCA algorithm for spectrum sensing is shown in Fig. \ref{fig:kpca}.
The detection with leading eigenvector under the framework of kernel PCA is simply called kernel PCA detection. The templates of PCA can be learned blindly even at very low signal to noise ratio (SNR) ~\cite{zhang2011spectrum}.

So far the mean of $\varphi ({\bf x}_i ),i = 1,2, \cdots M$ has been assumed to be zero. In fact, the zero mean data in the feature space are
\begin{equation}
\varphi ({\bf x}_i )-\frac{1}{M}\sum\limits_{i = 1}^M {\varphi ({\bf x}_i )}. 
\end{equation}
The kernel matrix for this centering or zero mean data can be derived by ~\cite{schölkopf1998nonlinear}
\begin{equation}
{{\bf K}_c} = {\bf K} - 1_M {\bf K} - {\bf K}1_M  + 1_M {\bf K}1_M 
\end{equation}
in which $(1_M )_{ij} : = 1/M$. The centering in feature space is not done in this paper.

Some commonly used kernels  are as follows:  polynomial kernels
\begin{equation}
\label{poly}
{\mathop{\rm k}\nolimits} ({\bf x}_i ,{\bf x}_j ) = (<{\bf x}_i  , {\bf x}_j>  + c)^{de}, c\geq0,
\end{equation}
where $de$ is the order of the polynomial,
radial basis kernels (RBF)
\begin{equation}
{\mathop{\rm k}\nolimits} ({\bf x}_i ,{\bf x}_j ) = \exp ( - \gamma \left\| {{\bf x}_i  - {\bf x}_j } \right\|^2 ),
\end{equation}
and Neural Network type kernels
\begin{equation}
{\mathop{\rm k}\nolimits} ({\bf x}_i ,{\bf x}_j ) = \tanh (<{\bf x}_i  , {\bf x}_j>  + b),
\end{equation}
in which the heavy-tailed RBF kernel is in the form of 
\begin{equation}
{\mathop{\rm k}\nolimits} ({\bf x}_i ,{\bf x}_j ) = \exp ( - \gamma \left\| {{\bf x}_i^a  - {\bf x}_j^a } \right\|^b ),
\end{equation}
and Gaussian RBF kernel is 
\begin{equation}
\label{rbf}
{\mathop{\rm k}\nolimits} ({\bf x}_i ,{\bf x}_j ) = \exp \left( { - \frac{{\left\| {{\bf x}_i  - {\bf x}_j } \right\|^2 }}{{2\sigma ^2 }}} \right).
\end{equation}

\section{Spectrum Sensing with GLRT and Kernel GLRT}
\label{kernelGLRT}

GLRT and kernel GLRT methods considered in this paper also assume that there is a training set ${\bf x}_1,{\bf x}_2, \cdots, {\bf   x}_M$ for the primary user's signal,  in which ${\bf x}_i,i=1,2,\cdots,M$ are $d-$dimensional column vectors.  The primary user's signal is assumed to lie on a given linear subspace  $[\bf {T }]$.  The training set is used to estimate this subspace $[\bf {T }]$.

Given the training set ${\bf x}_i,i=1,2,\cdots,M$, the sample covariance matrix ${\bf R}_x$ is obtained by \eqref{cov}. The eigenvectors of ${\bf R}_x$ corresponding to  nonzero eigenvalues are taken as the bases of the subspace $[\bf {T }]$.

Kernel GLRT ~\cite{kwon2006kernel}  based on matched subspace model for hyperspectral target detection has been proposed which takes into account the background. The background information can be taken as interference in spectrum sensing.  In this paper the modified kernel GLRT algorithm based on matched subspace model is proposed for spectrum sensing without taking into consideration the interference.

\subsection{GLRT Based on Matched Subspace Model}

The GLRT approach in this paper is based on the linear subspace model ~\cite{scharf1994matched} in which the primary user's signal is assumed to lie on a linear subspace  $[\bf {T }]$. Receiving one $d-$dimensional vector $\bf y$, the two hypotheses ${\mathcal H}_0$ and ${\mathcal H}_1$ can be expressed as
\begin{equation}
\begin{array}{l}
 {\mathcal H}_0 :{\bf y} = {\bf w} \\ 
 {\mathcal H}_1 :{\bf y} = {\bf {T }{\boldsymbol\theta} } + {\bf w}. \\ 
 \end{array}
\end{equation}
$[\bf {T }]$ is spanned by the column vectors of $\bf T$. $\bf T$ is an orthonormal matrix, ${\bf {T }}^{T}\bf {T } = \bf I$ in which $\bf I$ is an identity matrix. ${\boldsymbol\theta} $ is the coefficient's vector in which each entry representing the  magnitude  on each basis of $[\bf T]$. ${\bf w}$ is still white Gaussian noise vector which obeys multivariate Gaussian distribution $
N(0,\sigma ^2 {\bf I})$.

For the received vector ${\bf y}$, LRT approach  detects between the two hypotheses ${\mathcal H}_0$ and ${\mathcal H}_1$  by
\begin{equation}
\label{lrt}
\rho  = \frac{{f_1 ({\bf y}|{\mathcal H}_1 )}}{{f_0 ({\bf y}|{\mathcal H}_0 )}}\begin{array}{*{20}c}
   {{\mathcal H}_1 }  \\
   \gtrless \\
   {{\mathcal H}_0 }  \\
\end{array}T_{lrt}
\end{equation}
in which $T_{lrt}$ is the threshold value of LRT approach. $f_1 ({\bf y}|{\mathcal H}_1 )$ and $f_0 ({\bf y}|{\mathcal H}_0 )$ are conditional probability densities which follow Gaussian distributions,
\begin{equation}
\begin{array}{ll}
 {\mathcal H}_0& :f_0 ({\bf y}|{\mathcal H}_0 ): N(0,\sigma _0 ^2 {\bf I}) \\
 \\
 &= \frac{1}{{(2\pi \sigma _0 ^2 )^{d/2} }}\exp ( - \frac{1}{{2\sigma _0 ^2 }}\left\| {{\bf w}_0 } \right\|^2 ) ,\\ 
 \\ {\mathcal H}_1 &:f_1 ({\bf y}|{\mathcal H}_1 ): N({\bf T {\boldsymbol\theta} },\sigma _1 ^2 {\bf I}) \\
 \\
 &= \frac{1}{{(2\pi \sigma _1 ^2 )^{d/2} }}\exp ( - \frac{1}{{2\sigma _1 ^2 }}\left\| {{\bf w}_1 } \right\|^2 ) .\\ 

 \end{array}
\end{equation}

In general, the parameters ${\boldsymbol \theta },{\sigma _0},{\sigma _1}$ are unknown to us under which the GLRT approach is explored. In GLRT, the parameters
 ${\boldsymbol \theta},\sigma _0,\sigma _1$ are replaced by their maximum likelihood estimates ${\boldsymbol {\hat \theta }},{\hat {\sigma }_0},{\hat {\sigma} _1} $.
The maximum likelihood estimate of ${\boldsymbol \theta } $ is equivalent to the least square estimate of ${\bf {w}}_1$ ~\cite{kwon2006kernel},
\begin{equation}
\begin{array}{l}
 {\bf \hat w}_0  = {\bf y} \\ 
 {\bf \hat w}_1  = {\bf y} - {\bf {T} \hat{ \boldsymbol \theta }} = ({\bf I} - P_{\bf T}){\bf y}. \\ 
\end{array}
\end{equation}
${\hat {\sigma }_0},{\hat {\sigma} _1} $ can be cast as 
\begin{equation}
\begin{array}{l}
\hat \sigma _0 ^2  = \frac{1}{d}\left\| {{\bf \hat w}_0 } \right\|^2 \\
\hat \sigma _1 ^2  = \frac{1}{d}\left\| {{\bf \hat w}_1 } \right\|^2 .\\
\end{array}
\end{equation}

Substituting the maximum likelihood estimates of the parameters into \eqref{lrt} and taking $d/2$ root, GLRT is expressed as ~\cite{scharf1994matched}
\begin{equation}
\begin{array}{ll}
 \rho  = \frac{{\left\| {{\bf \hat w}_0 } \right\|^2 }}{{\left\| {{\bf \hat w}_1 } \right\|^2 }} &= \frac{{{\bf y}^T P_{\bf I} {\bf y}}}{{{\bf y}^T (P_{\bf I} - P_{\bf T} ){\bf y}}} \\ 
  &= \frac{{{\bf y}^T {\bf y}}}{{{\bf y}^T ({\bf I} - {\bf TT}^T ){\bf y}}} \\ 
 \end{array}
\end{equation}
where $ P_{\bf I} =\bf I$ is the identity projection operator, and $P_{\bf T}$ is  the projection  onto the subspace $[\bf {T }]$,
\begin{equation}
\begin{array}{l}
 P_{\bf T}  = {\bf T}({\bf T}^T {\bf T})^{ - 1} {\bf T}{}^T = {\bf TT}{}^T .\\ 
 \end{array}
\end{equation}
The detection result is evaluated by comparing $\rho$ of GLRT with a threshold value  $T_{glrt}$.

\subsection{Kernel GLRT Based on Matched Subspace Model}

Accordingly, if ${\mathcal H}_{0_\varphi}$, ${\mathcal H}_{1_\varphi}$ also obey Gaussian distributions ~\cite{kwon2006kernel}
\begin{equation}
\begin{array}{l}
 {\mathcal H}_{0_\varphi} :\varphi ({\bf y}) = {\bf w}_\varphi   \\ 
 {\mathcal H}_{1_\varphi} :\varphi ({\bf y}) = {\bf T}_\varphi  {\boldsymbol \theta }_\varphi   + {\bf w}_\varphi ,  \\ 
 \end{array}
\end{equation}
then GLRT can be  extended to the feature space of ${\varphi ({\bf y})}$,
\begin{equation}
\label{GLRT}
\rho  = \frac{{\left\| {{\bf \hat w}_{0_\varphi  } } \right\|^2 }}{{\left\| {{\bf \hat w}_{1_\varphi  } } \right\|^2 }} = \frac{{{\varphi ({\bf y})}^T P_{{\bf I}_\varphi  } {\varphi ({\bf y})}}}{{{\varphi ({\bf y})}^T (P_{{\bf I}_\varphi  }  - P_{{\bf T}_\varphi  } ){\varphi ({\bf y})}}}
\end{equation}
where ${P_{{\bf I}_\varphi  } }$ is the identity projection operator in the feature space. $[{\bf T}_\varphi]$ is the linear space that the primary user's signal in the feature space lies on.  Each column of ${\bf T}_\varphi$ is the eigenvector corresponding to the nonzero eigenvalue of 
\begin{equation}
{\bf R}_{\varphi (x)}  =  \frac{1}{M} \sum\limits_{i = 1}^M {\varphi ({\bf x}_i )} \varphi ({\bf x}_i )^T .
\end{equation}

Likewise, $P_{{\bf T}_\varphi  }$ is a projection operator onto the primary signal's subspace,
\begin{equation} 
P_{{\bf T}_\varphi  }  = {\bf T}_\varphi  ({\bf T}_\varphi  ^T {\bf T}_\varphi  )^{ - 1} {\bf T}_\varphi  {}^T = {\bf T}_\varphi  {\bf T}_\varphi  {}^T.
\end{equation}

Here, we assume that  ${P_{{\bf I}_\varphi  } }$ can  perfectly project ${\varphi ({\bf x})}$ as  ${\varphi ({\bf x})}$  in the feature space which is different from the method proposed in  ~\cite{kwon2006kernel},
\begin{equation}
{\varphi ({\bf y})^T P_{{\bf I}_\varphi  } \varphi ({\bf y})}={\varphi ({\bf y})^T  \varphi ({\bf y})}.
\end{equation}

Based on the derivation of kernel PCA, the eigenvectors corresponding to  nonzero eigenvalues of  the sample covariance matrix  ${\bf R}_ {\varphi(x)}$  are $(\varphi ({\bf x}_1 ),\varphi ({\bf x}_2 ),...,\varphi ({\bf x}_M ))({\boldsymbol\beta} _1 ,{\boldsymbol\beta} _2 ,...,{\boldsymbol\beta} _K )$. 
${\boldsymbol\beta} _1 ,{\boldsymbol\beta} _2 ,...,{\boldsymbol\beta} _K $ are eigenvectors  corresponding to nonzero eigenvalues of
 ${\bf K} = ({\mathop{\rm k}\nolimits} ({\bf x}_i ,{\bf x}_j ))_{ij}$. $K$ is the number of nonzero eigenvalues of $\bf K$. 
Accordingly, ${\varphi ({\bf y})^T P_{{\bf T}_\varphi  } \varphi ({\bf y})}$ can be represented as  

\begin{equation}
\label{mid}
\begin{array}{l}
 \varphi ({\bf y})^T {\bf T}_\varphi  {\bf T}_\varphi  {}^T\varphi ({\bf y}) = \varphi ({\bf y})^T (\varphi ({\bf x}_1 ),\varphi ({\bf x}_2 ),...,\varphi ({\bf x}_M )) \cdot  \\ 
 \\
 ({\boldsymbol\beta} _1 ,{\boldsymbol\beta} _2 ,...,{\boldsymbol\beta} _K )({\boldsymbol\beta} _1 ,{\boldsymbol\beta} _2 ,...,{\boldsymbol\beta} _K )^T \left( \begin{array}{c}
 \varphi ({\bf x}_1 )^T  \\ 
 \varphi ({\bf x}_2 )^T  \\ 
  \vdots  \\ 
 \varphi ({\bf x}_M )^T  \\ 
 \end{array} \right)\varphi ({\bf y}) \\ 
 \\
  = ({\mathop{\rm k}\nolimits} ({\bf y},{\bf x}_1 ),{\mathop{\rm k}\nolimits} ({\bf y},{\bf x}_2 ),...,{\mathop{\rm k}\nolimits} ({\bf y},{\bf x}_M ))({\boldsymbol\beta} _1 ,{\boldsymbol\beta} _2 ,...,{\boldsymbol\beta} _K ) \cdot  \\ 
  \\
 ({\boldsymbol\beta} _1 ,{\boldsymbol\beta} _2 ,...,{\boldsymbol\beta}_K )^T \left( \begin{array}{c}
 {\mathop{\rm k}\nolimits} ({\bf y},{\bf x}_1 ) \\ 
 {\mathop{\rm k}\nolimits} ({\bf y},{\bf x}_2 ) \\ 
  \vdots  \\ 
 {\mathop{\rm k}\nolimits} ({\bf y},{\bf x}_M ) \\ 
 \end{array} \right) .\\ 
 \end{array}
\end{equation}

The derivation of \eqref{GLRT} is based on the assumption that the hypotheses ${\mathcal H}_{0_\varphi}$,${\mathcal H}_{1_\varphi}$  obey Gaussian distributions. The paper ~\cite{kwon2006kernel} has claimed that, though without strict proof, if ${\mathop{\rm k}}$ is Gaussian kernel ${\mathcal H}_{0_\varphi}$,${\mathcal H}_{1_\varphi}$  are still distributed as Gaussian's. 

 Gaussian kernel is employed for the kernel GLRT approach, thus $\varphi ({\bf y})^T \varphi ({\bf y}) = {\mathop{\rm k}\nolimits} ({\bf y},{\bf y}) = 1$. Substituting \eqref{mid} into \eqref{GLRT}, 
\begin{equation}
\label{kglrt}
\begin{array}{l}
 \rho  = \frac{1}{{1 - {\bf k}_{\bf T} ^T ({\boldsymbol\beta} _1 ,{\boldsymbol\beta} _2 ,...,{\boldsymbol\beta} _K )\left( \begin{array}{c}
 {\boldsymbol\beta} _1 ^T  \\ 
 {\boldsymbol\beta} _2 ^T  \\ 
  \vdots  \\ 
 {\boldsymbol\beta} _K ^T  \\ 
 \end{array} \right){\bf k}_{\bf T} }} \\ 
 \end{array}
\end{equation}
in which 
\begin{equation}
\label{test}
{{\bf k}_{\bf T}  = \left( \begin{array}{c}
 k({\bf y},{\bf x}_1 ) \\ 
 k({\bf y},{\bf x}_2 ) \\ 
  \vdots  \\ 
 k({\bf y},{\bf x}_M ) \\ 
 \end{array} \right)}.
\end{equation}

The centering of ${{\bf k}_{\bf T} }$ in the feature space  ~\cite{kwon2006kernel} is 
\begin{equation}
{\bf k}_{\bf T}  = {\bf k}_{\bf T}  - \left( \begin{array}{l}
 1 \\ 
 1 \\ 
  \vdots  \\ 
 1 \\ 
 \end{array} \right)\left( {\frac{1}{M},\frac{1}{M},...,\frac{1}{M}} \right){\bf k}_{\bf T} .
\end{equation}

\begin{figure}[!t]
\begin{center}
\scalebox{.40}{\includegraphics{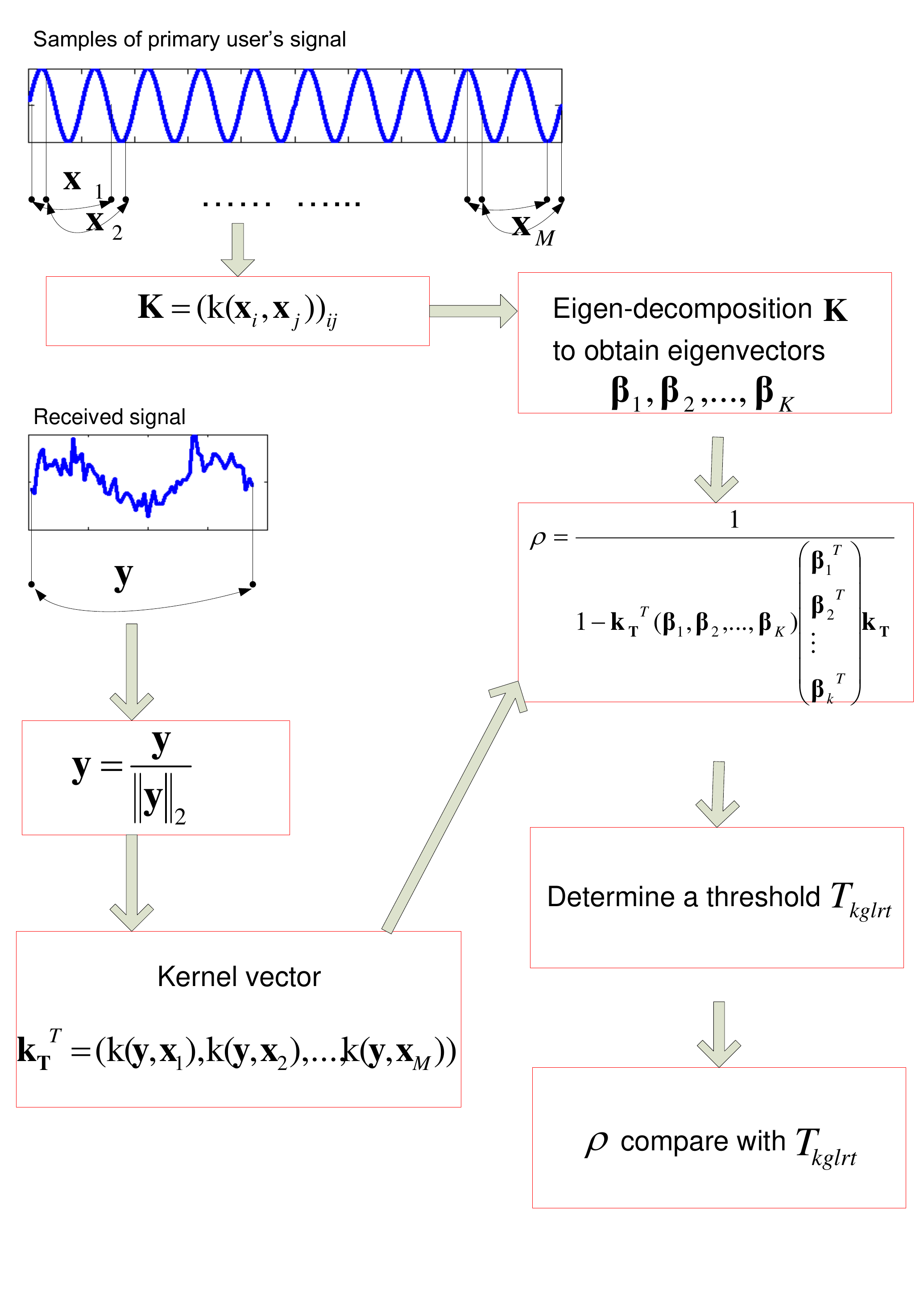}}
\end{center}
\vspace{-1ex}
\caption{The flow chart of the proposed kernel GLRT algorithm for spectrum sensing}
\vspace{-2ex}
\label{fig:kglrt}
\end{figure}

The procedure of kernel GLRT  for spectrum sensing based on Gaussian kernels without consideration of centering is summarized here as follows:
\begin{enumerate}
\item Given a training set of the primary user's signal ${\bf x}_1,{\bf x}_2,\cdots,{\bf x}_M$, the kernel matrix is ${\bf K} = ({\mathop{\rm k}\nolimits} ({\bf x}_i ,{\bf x}_j ))_{ij}$.
 $\bf K$ is positive semidefinite. Eigen-decomposition of $\bf K$ to obtain eigenvectors ${\boldsymbol \beta }_{1 },{\boldsymbol \beta }_{2 },\cdots, {\boldsymbol \beta }_{K }$ corresponding to all of the nonzero eigenvalues.
\item Normalize the received $d-$dimensional vector $\bf y$ by
\begin{equation}
{\bf y} = \frac{{\bf y}}{{\left\| {\bf y} \right\|_2 }}.
\end{equation}
\item Compute the kernel vector of ${{\bf k}_{\bf T}}$ by \eqref{test}.
\item Compute the value of $\rho$ defined in \eqref{kglrt}.
\item Determine a threshold value $T_{kglrt}$ for a desired false alarm rate.
\item Detect the presence or absence of ${\bf x} $ in ${\bf y} $ by checking $\rho>T_{kglrt}$ or not.
\end{enumerate}
The flow chart of the proposed kernel GLRT algorithm for spectrum sensing with Gaussian kernels is shown in Fig. \ref{fig:kglrt}.

The detection rate and false alarm rate for all of the above methods can be calculated by
\begin{equation}
\begin{array}{l}
\begin{array}{l}
 P_d  =  {prob}(\rho  > T|{\bf y} = {\bf x} + {\bf w}) \\ 
 P_f  =  {prob}(\rho  > T|{\bf y} = {\bf w}) \\ 
 \end{array}
 \end{array}
\end{equation}
where $T$ is the threshold value determined by each of the above algorithm. In general,  threshold value is determined by  false alarm rate of $10\%$.

\section{Experiments}
\label{experiments}

The experimental results will be compared with the results of estimator-correlator (EC) ~\cite{steven1998fundamentals} and maximum minimum eigenvalue (MME) ~\cite{zeng2006maximum}.
EC method assumes that the signal $\bf x$ follows zero mean Gaussian distribution with the  covariance matrix ${\bf \Sigma }_x$, 
\begin{equation}
{\bf x}:N(0,{\bf \Sigma }_x ),\:\:\: {\bf w}:N(0,\sigma ^2 {\bf I}).
\end{equation}
Both ${\bf \Sigma }_x$ and $\sigma ^2$ are given priorly. Consequently, when signal $\bf x$ obeys Gaussian distribution, EC method is optimal. The hypothesis is ${\mathcal H}_1$ when
\begin{equation}
\label{ec}
\rho  = {\bf y}^T {\bf \Sigma }_x ({\bf \Sigma }_x  + \sigma ^2 {\bf I})^{ - 1} {\bf y} > T_{ec},
\end{equation}
 where $T_{ec}$ is the threshold value designed for the EC method. 
 
MME is a totally blind method without any prior knowledge on the covariance matrix of the signal and $\sigma ^2$. The hypothesis is ${\mathcal H}_1$ when
\begin{equation}
\frac{{\tilde \lambda _{\max } }}{\tilde \lambda _{\min } } > T_{mme},
\end{equation}
 where $T_{mme}$ is the threshold value designed for the MME method. ${\tilde \lambda _{\max }  }$ and ${\tilde \lambda _{\min }}$ are the maximal and minimal eigenvalues of the sample covariance matrix ${\bf R}_y= \frac{1}{M}\sum\limits_{i = 1}^M {{\bf y}_i } {\bf y}_{i }^T$. 
 
PCA, kernel PCA, GLRT, and kernel GLRT methods considered in this paper bear partial prior knowledge, that is, the sample covariance matrix of the signal is given priorly. 

\subsection{Experiments on the Simulated Sinusoidal Signal}

The primary user's signal assumes to be the sum of three sinusoidal functions with unit amplitude of each. The generated sinusoidal samples with length $L =500$  are taken as the samples of  $x(n)$. The training set ${\bf x}_1,{\bf x}_2, \cdots, {\bf   x}_M$ is taken from $x(n)$ with $d =128$ and $i=1$. Received signal $y(n)$ is the same length as $x(n)$. Vectorized $y(n)$ are ${\bf y}_1,{\bf y}_2, \cdots, {\bf y}_M$ with $d =128$ and $i=1$.  For the received vectors ${\bf y}_1,{\bf y}_2, \cdots, {\bf y}_M$, EC detection is implemented on every vector and then do average (same implementation for GLRT and kernel GLRT)
\begin{equation}
\rho  = \frac{1}{M}\sum\limits_{i = 1}^M {{\bf y}_i ^T {\bf \Sigma }_x } ({\bf \Sigma }_x  + \sigma ^2 {\bf I})^{-1}{\bf y}_i.
\end{equation}

Polynomial kernel of order 2 with $c=1$ is applied for kernel PCA. 

\begin{figure}[!t]
\begin{center}
\scalebox{.40}{\includegraphics{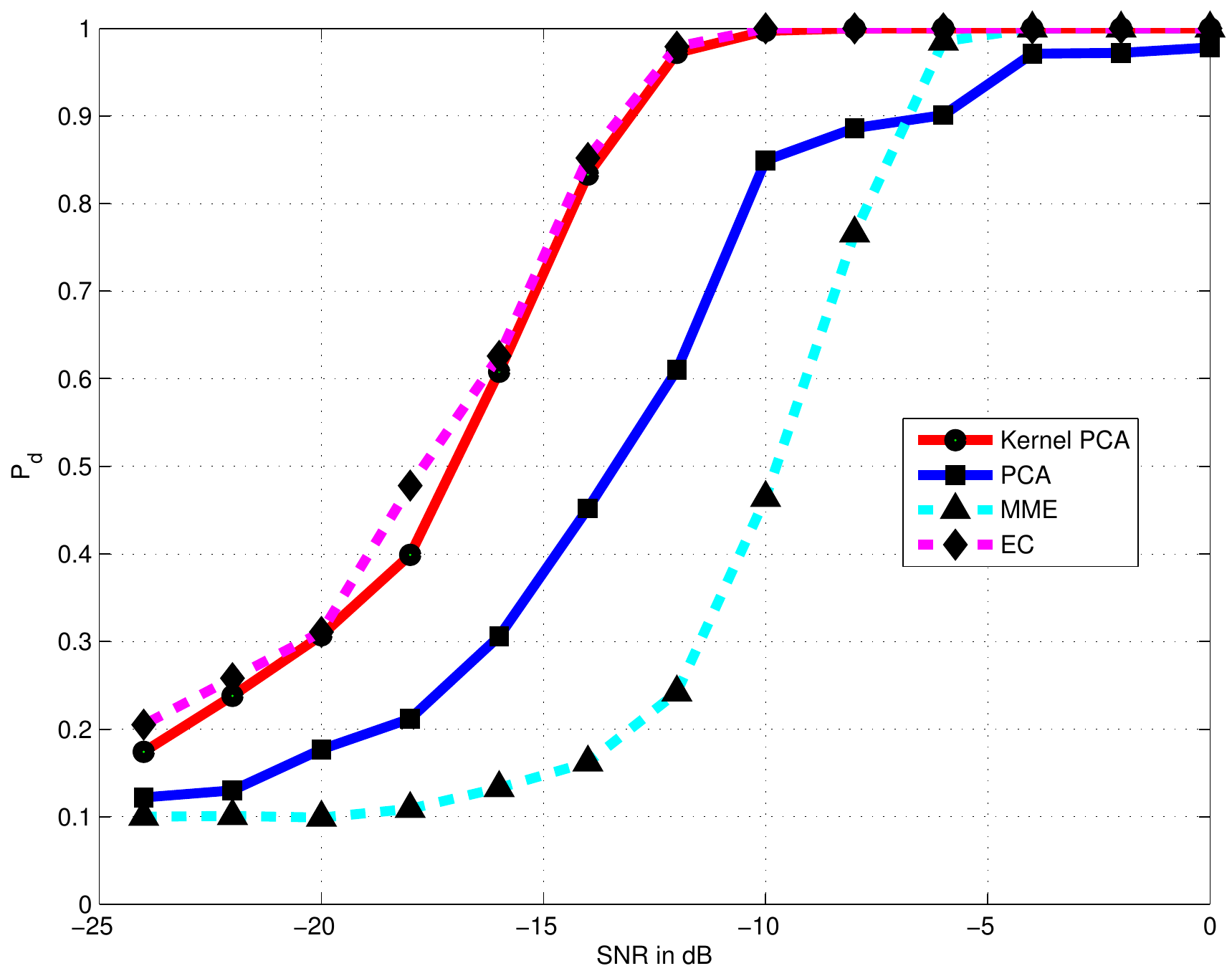}}
\end{center}
\vspace{-1ex}
\caption{The detection rates for kernel PCA and PCA compared with EC and MME with $P_f =10\%$ for the simulated signal}
\vspace{-2ex}
\label{fig:KPCA_pd}
\end{figure}

The detection rates varied by SNR for kernel PCA and PCA compared with EC and MME with $P_f =10\%$ are shown in Fig. \ref{fig:KPCA_pd} for 1000 experiments. 
From Fig. \ref{fig:KPCA_pd}, it can be seen that when SNR $<$ -10 dB, kernel PCA is about 4 dB better than PCA method. Kernel PCA can compete with EC method but with less known prior knowledge. It should be noticed that the types of kernel functions and parameters in kernel functions can both affect the performance of the kernel PCA approach.  

\begin{figure}[!t]
\begin{center}
\scalebox{.40}{\includegraphics{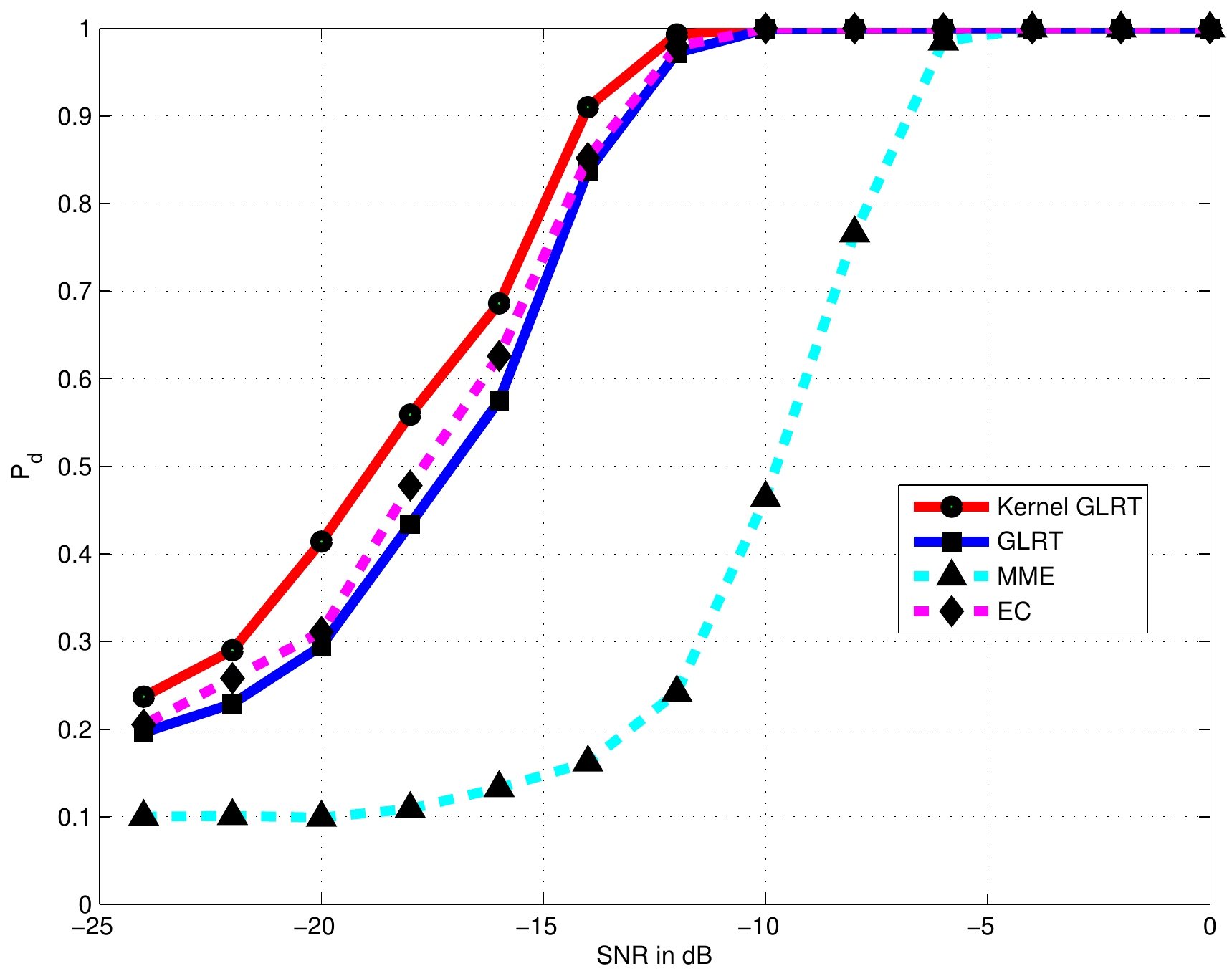}}
\end{center}
\vspace{-1ex}
\caption{The detection rates for kernel GLRT  and GLRT compared with EC and MME with $P_f =10\%$ for the simulated signal}
\vspace{-2ex}
\label{fig:KGLRT_pd}
\end{figure}

The detection rates varied by SNR for kernel GLRT and GLRT compared with EC and MME with $P_f =10\%$ are shown in Fig. \ref{fig:KGLRT_pd} for 1000 experiments. Kernel GLRT is  still better than GLRT method.  Kernel GLRT can even beat EC method. The underlying reason is that EC method  assumes sinusoidal signal also following zero-mean Gaussian distribution with the actual distribution of which being shown in Fig. \ref{fig:distribution}. As is well known that sinusioal signal lies on a linear subspace which can be nearly perfectly estimated from the sample covariance matrix. Therefore, the matched subspace model for GLRT and kernel GLRT considered in this paper is more suitable for sinusoial signal.  Gaussian kernel is used with the parameter $\sigma  = \frac{{15}}{{\sqrt 2 }}$. The width of Gaussian kernel $\sigma $ is the major factor that affects the performance of the kernel GLRT approach.

\begin{figure}[!t]
\begin{center}
\scalebox{.40}{\includegraphics{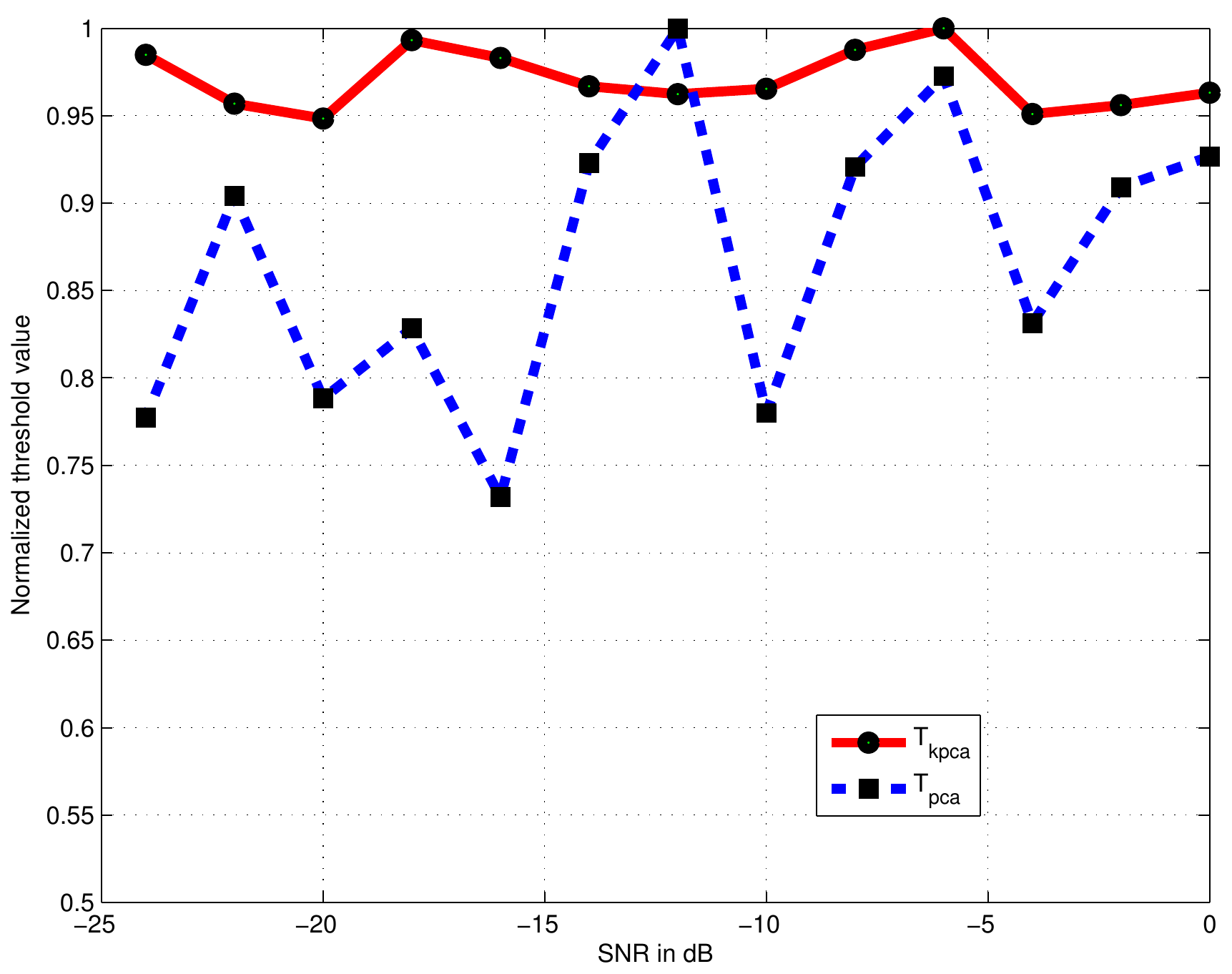}}
\end{center}
\vspace{-1ex}
\caption{Normalized threshold values for kernel PCA and PCA }
\vspace{-2ex}
\label{fig:KPCA_threshold}
\end{figure}

\begin{figure}[!t]
\begin{center}
\scalebox{.40}{\includegraphics{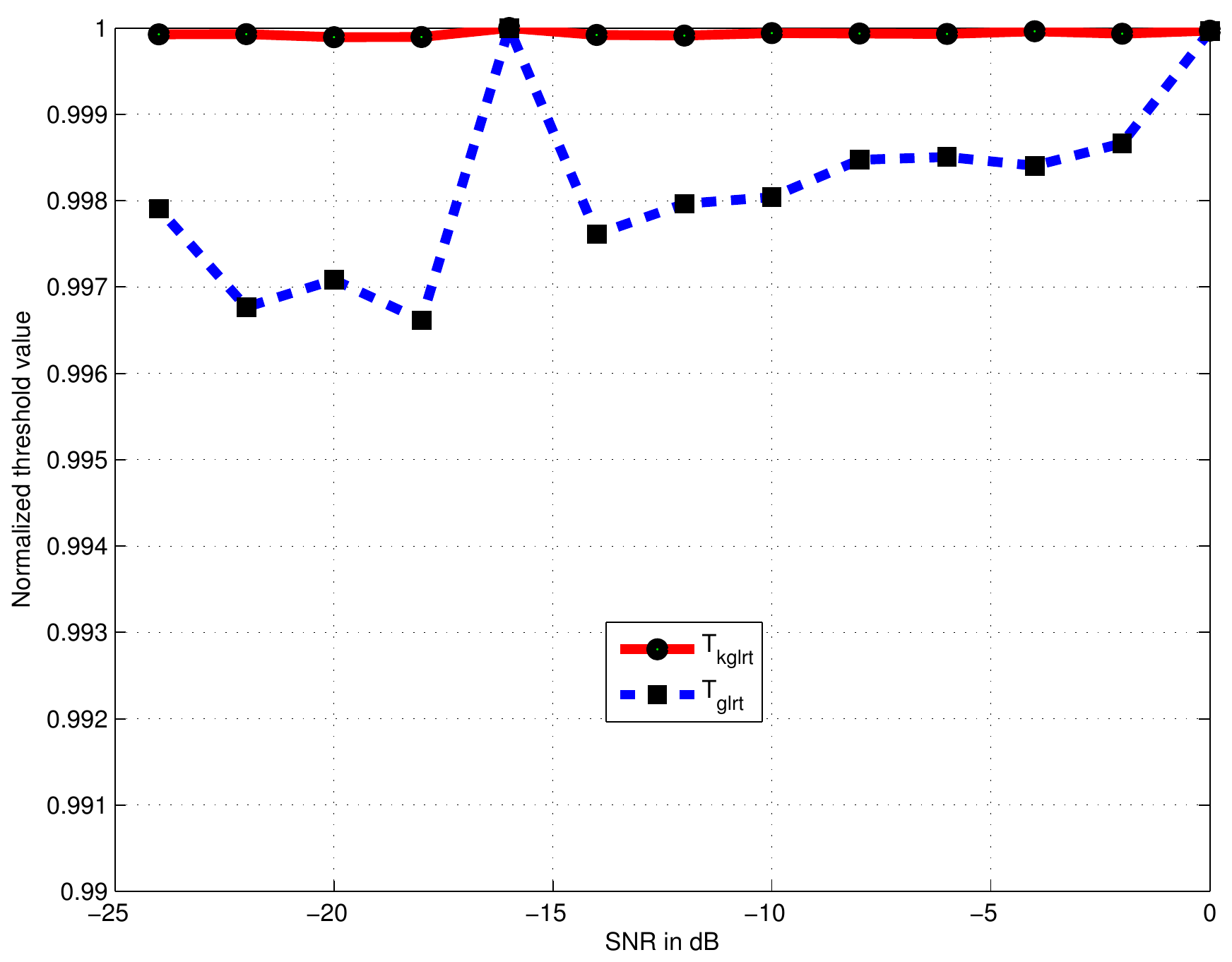}}
\end{center}
\vspace{-1ex}
\caption{Normalized threshold values for kernel GLRT and GLRT }
\vspace{-2ex}
\label{fig:KGLRT_threshold}
\end{figure}

The calculated threshold values with $P_f =10\%$ for kernel PCA, PCA,  kernel GLRT, and GLRT methods  are shown in  Fig. \ref{fig:KPCA_threshold} and Fig. \ref{fig:KGLRT_threshold}, respectively.  The threshold values are normalized by dividing the corresponding  maximal values  in $T_{pca}$, $T_{kpca}$, $T_{glrt}$ and $ T_{kglrt}$, respectively.
The threshold values assigned for the kernel methods are more stable than the corresponding linear methods. 

The simulation results are tested by choosing the kernel function as ${\mathop{\rm k}\nolimits} ({\bf x}_{i} ,{\bf x}_j ) = <{\bf x}_{i} ,{\bf x}_j > $. In this manner, the selected feature space is the original space. If the operations in the feature space and original space are identical, (for example, the centering is done in both of the spaces, and similarity measure is the inner-product for both PCA and kernel PCA), the results for kernel and corresponding linear methods should be the same. The tested results verified the correctness of the simulation.

%
%

\subsection{Experiments on Captured DTV Signal}

DTV signal ~\cite{Tawil} captured in Washington D.C. will be employed to the experiment of spectrum sensing in this section. The first segment of DTV signal with $L=500$ is taken as the samples of the primary user's signal $x(n)$.

\begin{figure}[!t]
\begin{center}
\scalebox{.40}{\includegraphics{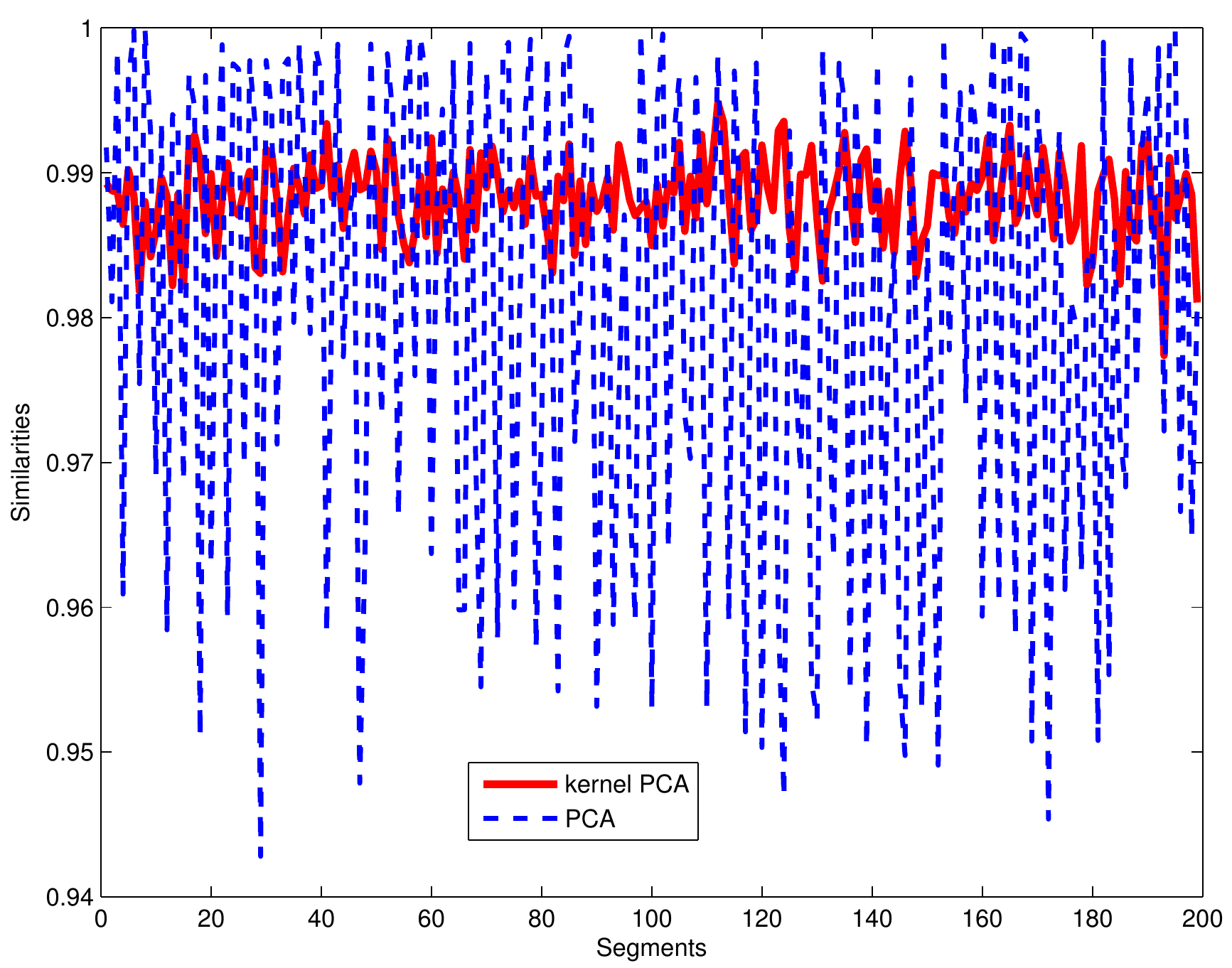}}
\end{center}
\vspace{-1ex}
\caption{Similarities of leading eigenvectors derived by PCA and kernel PCA  between the first segment and other $199$ segments}
\vspace{-2ex}
\label{fig:simi_DTV}
\end{figure}

First, the similarities  of leading eigenvectors of the sample covariance matrix between first segment and other segments  of DTV signal will be tested under the frameworks of PCA and kernel PCA. The DTV signal with length $10^5$ is obtained and divided into $200$ segments with the length of each segment $500$. Similarities of leading eigenvectors derived by PCA and kernel PCA  between the first segment and the rest $199$ segments are shown in Fig. \ref{fig:simi_DTV}. The result shows that the similarities are very high between leading eigenvectors of different segment's DTV signal (which are all above $0.94$), on the other hand, kernel PCA is more stable than PCA. 

\begin{figure}[!t]
\begin{center}
\scalebox{.40}{\includegraphics{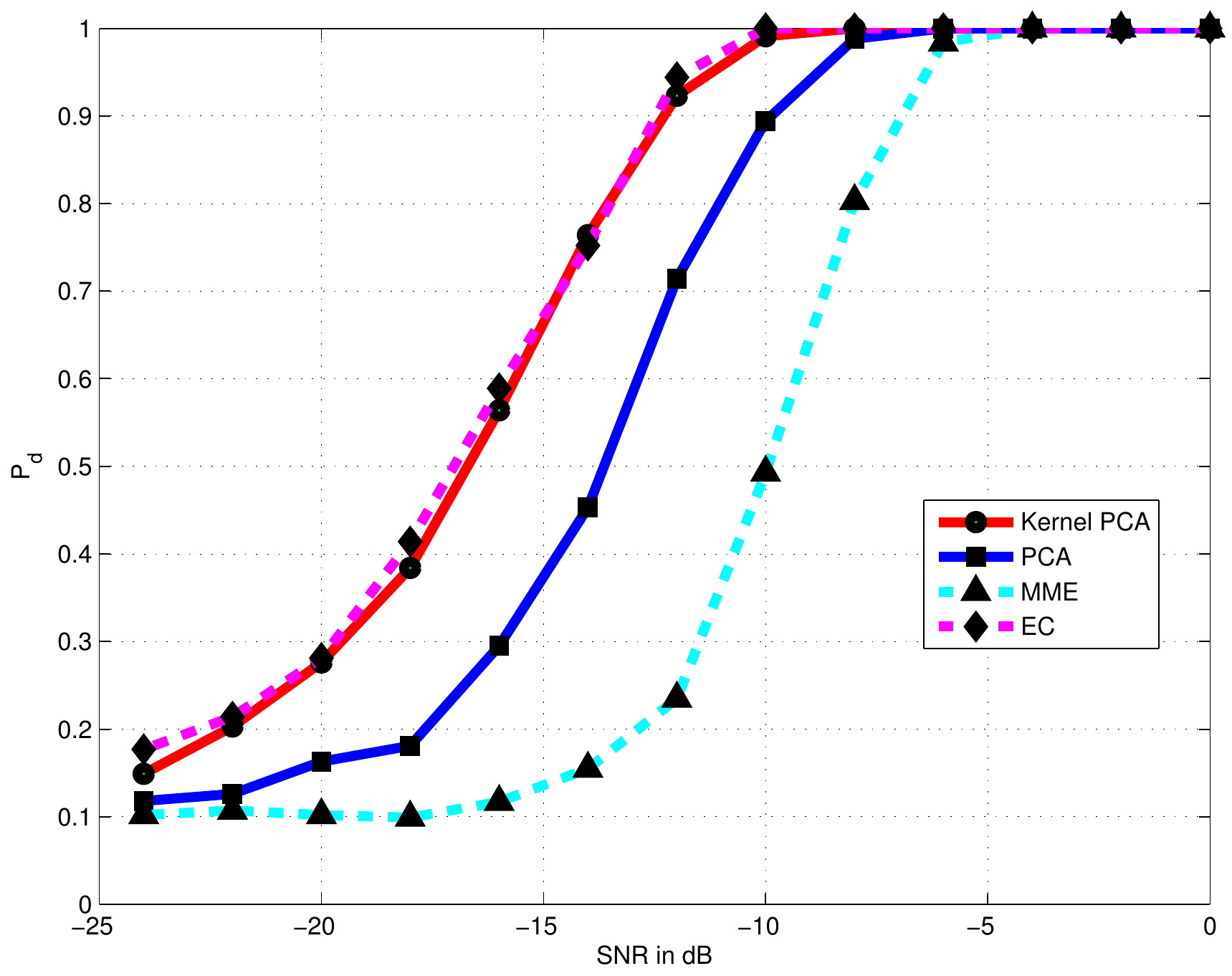}}
\end{center}
\vspace{-1ex}
\caption{The detection rates for kernel PCA  and PCA compared with EC and MME with $P_f =10\%$ for DTV signal }
\vspace{-2ex}
\label{fig:KPCA_pd_DTV}
\end{figure}

\begin{figure}[!t]
\begin{center}
\scalebox{.40}{\includegraphics{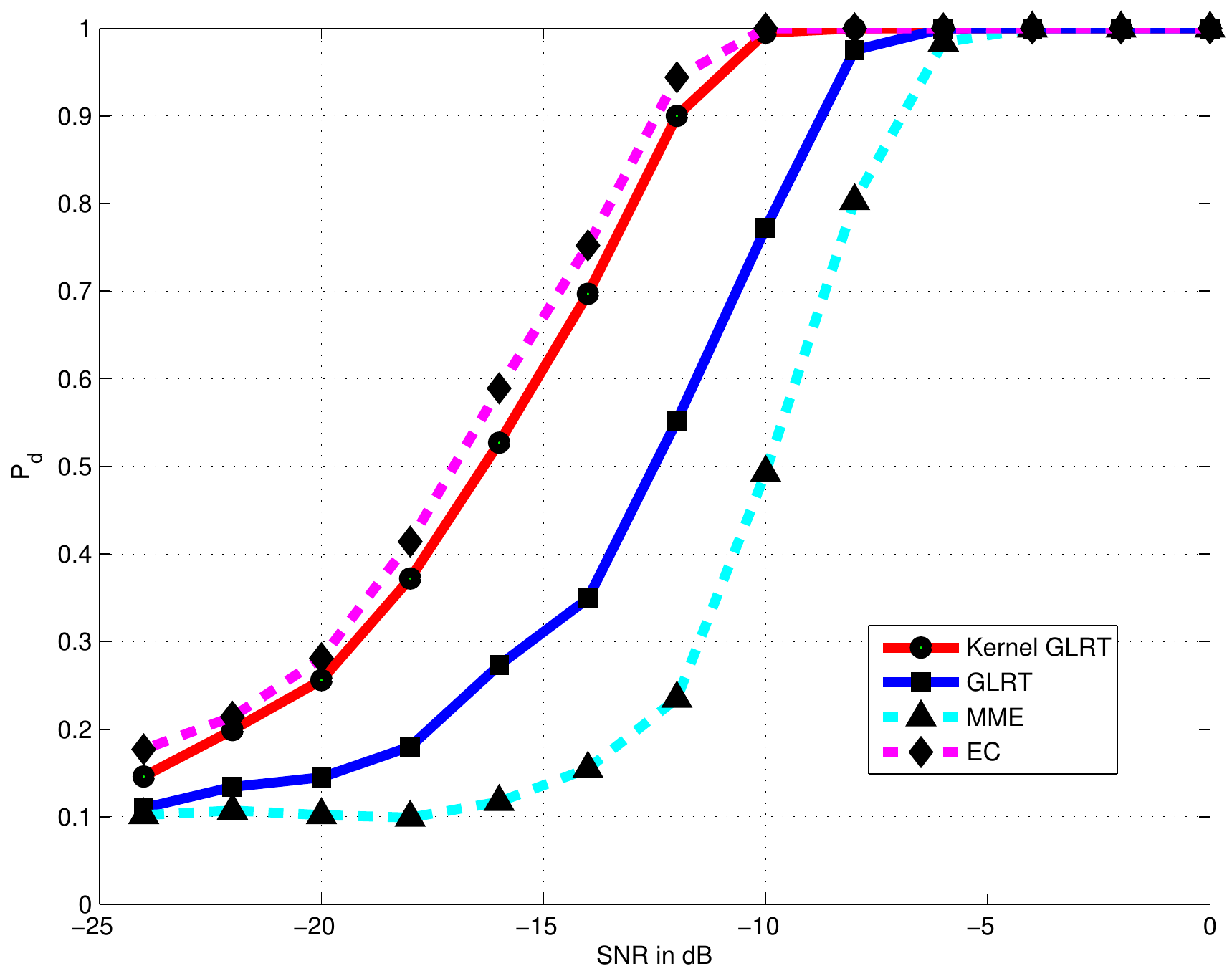}}
\end{center}
\vspace{-1ex}
\caption{The detection rates for kernel GLRT  and GLRT compared with EC and MME with $P_f =10\%$ for DTV signal }
\vspace{-2ex}
\label{fig:KGLRT_pd_DTV}
\end{figure}

\begin{figure}[!t]
\begin{center}
\scalebox{.40}{\includegraphics{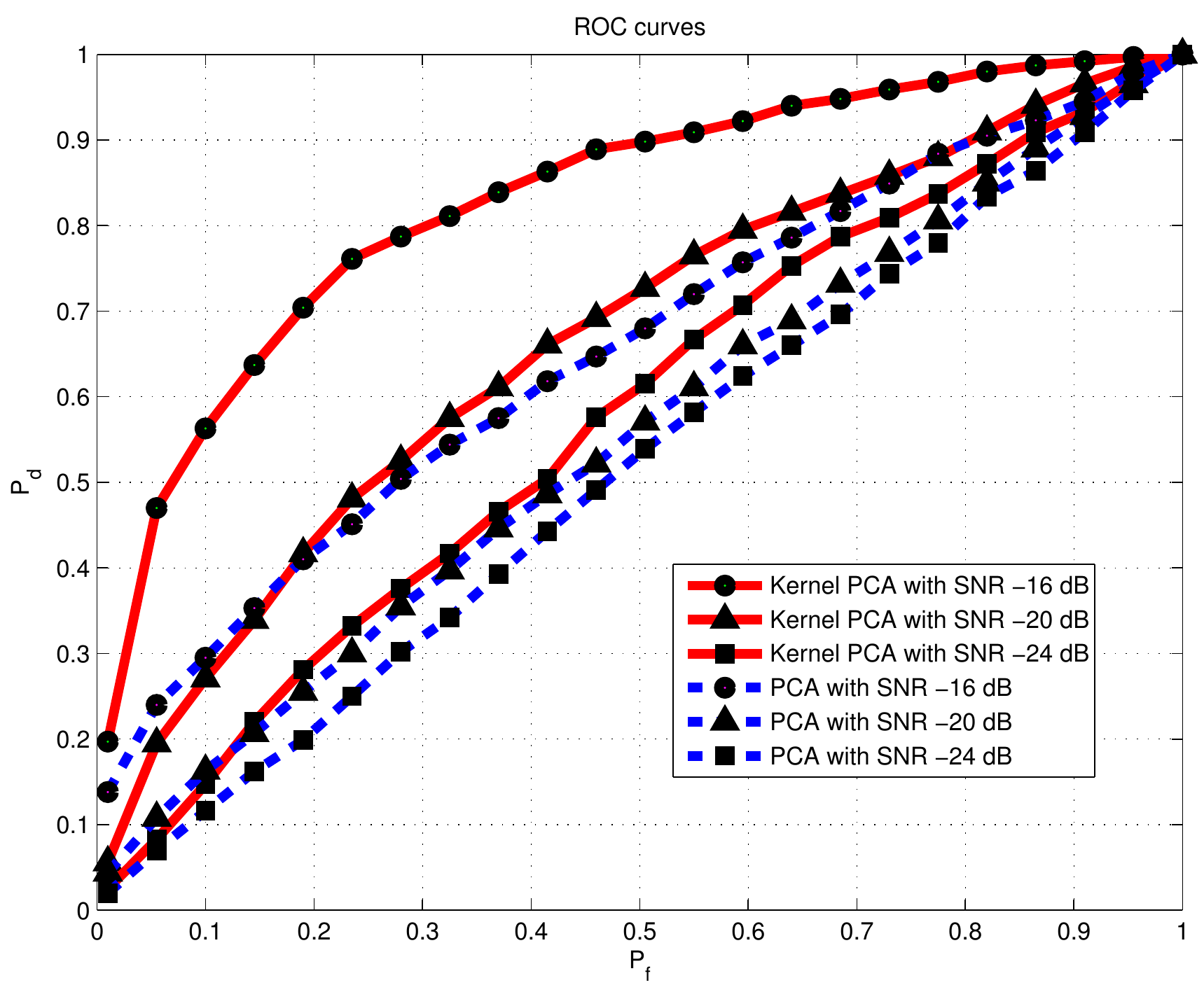}}
\end{center}
\vspace{-1ex}
\caption{ROC curves for kernel PCA and PCA for DTV signal}
\vspace{-2ex}
\label{fig:KPCA_roc_DTV}
\end{figure}

\begin{figure}[!t]
\begin{center}
\scalebox{.40}{\includegraphics{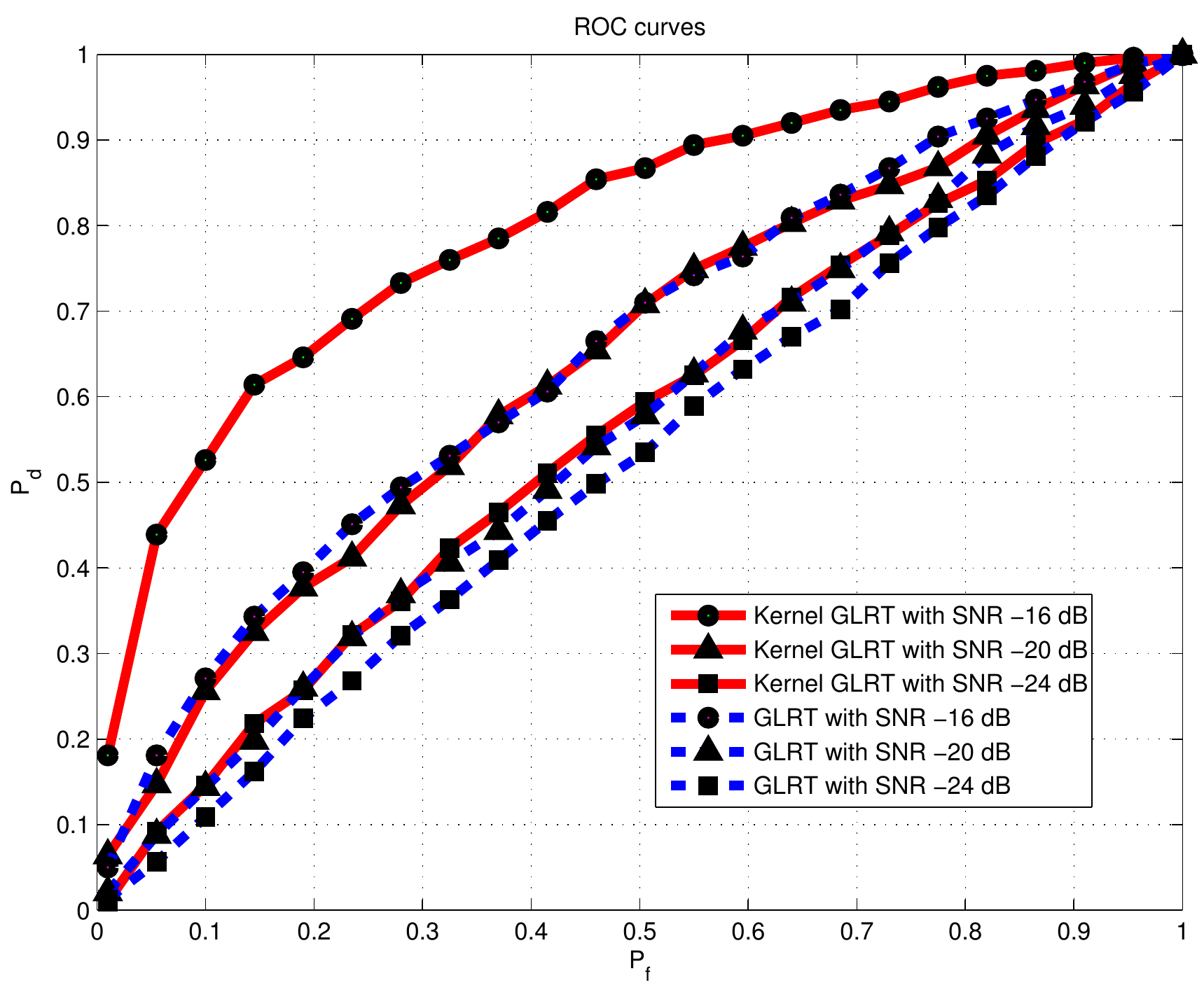}}
\end{center}
\vspace{-1ex}
\caption{ROC curves for kernel GLRT and GLRT for DTV signal }
\vspace{-2ex}
\label{fig:KGLRT_roc_DTV}
\end{figure}

\begin{figure}[!t]
\begin{center}
\scalebox{.45}{\includegraphics{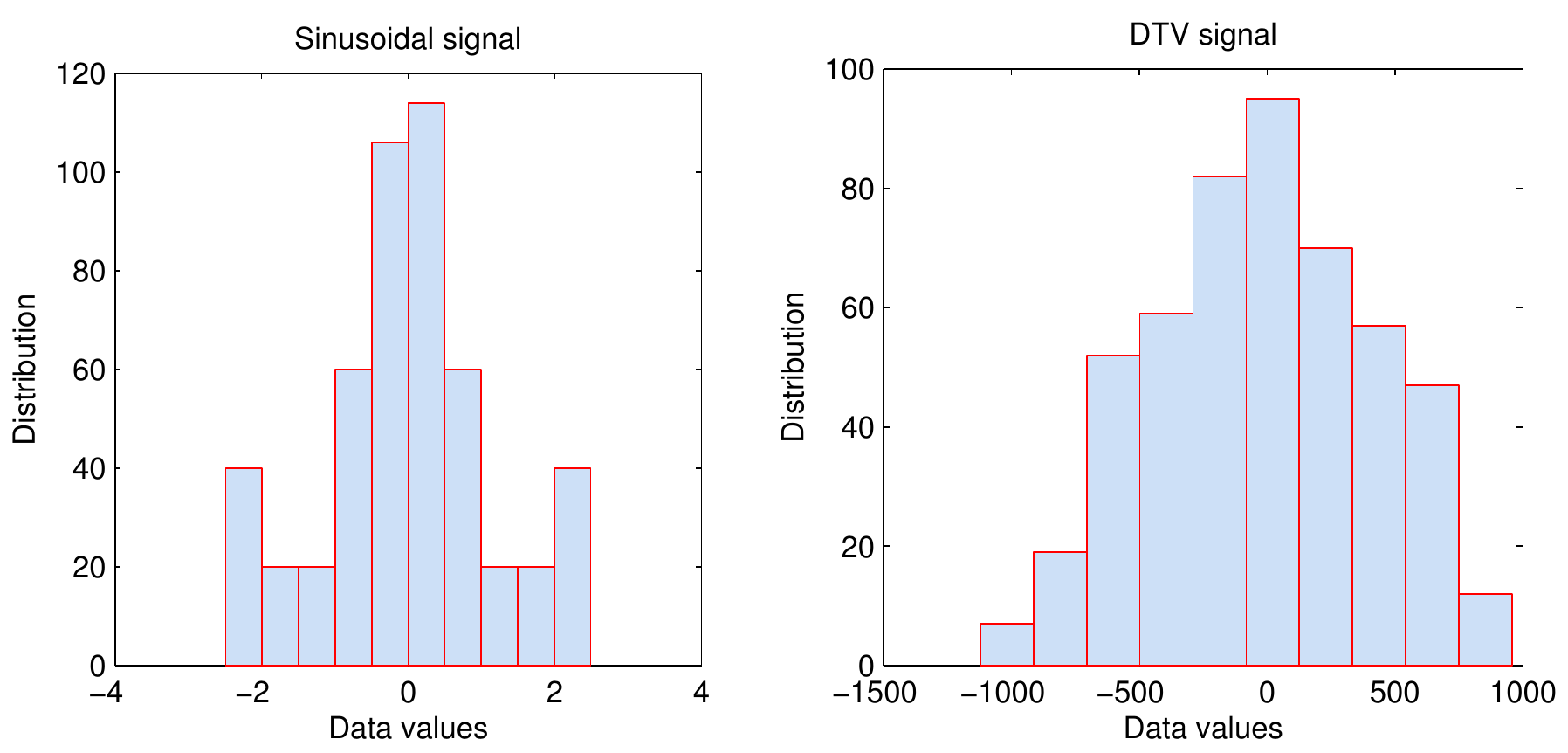}}
\end{center}
\vspace{-1ex}
\caption{The histograms of sinusoidal and DTV signal }
\vspace{-2ex}
\label{fig:distribution}
\end{figure}

The detection rates varied by SNR for kernel PCA and PCA (kernel GLRT and GLRT) compared with EC and MME with $P_f =10\%$ are shown in Fig. \ref{fig:KPCA_pd_DTV} (Fig. \ref{fig:KGLRT_pd_DTV}) for 1000 experiments. The  ROC curves are shown in Fig. \ref{fig:KPCA_roc_DTV} ( Fig. \ref{fig:KGLRT_roc_DTV}) for kernel PCA  and PCA (kernel GLRT and GLRT) with SNR = -16, -20, -24 dB. Experimental results show that kernel methods are 4 dB better than the corresponding linear methods. Kernel methods can compete with EC method. Howerver, kernel GLRT in this example cannot beat EC method due to the fact that the distribution of DTV signal (shown in  Fig. \ref{fig:distribution}) is more approximated Gaussian than the above simulated sinusoidal signal. 
Gaussian kernel with  parameter $\sigma  = \frac{{0.5}}{{\sqrt 2 }}$ is applied for kernel GLRT. Polynomial kernel of order 2 with $c=1$ is applied for kernel PCA.

\section{Conclusion}
\label{conclusion}

Kernel methods have been extensively and effectively applied in machine learning. Kernel is a very powerful tool in machine learning. Kernel function can extend the linear method to nonlinear one by defining the inner-product of data in the feature space.  The mapping from the original space to a higher dimensional feature space is indirectly defined by the kernel function. Kernel method makes the computation in an arbitrary dimensional feature space become possible.

In this paper, the detection with the leading eigenvector under the framework of kernel PCA is proposed. The inner-product between leading eigenvectors is taken as the similarity measure for kernel PCA approach. The proposed algorithm makes the detection in  an arbitrary dimensional feature space become possible. Kernel GLRT based on matched subspace model is also  introduced to spectrum sensing. Different from ~\cite{kwon2006kernel}, the kernel GLRT approach proposed in this paper assumes that identity  projection operator $P_{{\bf I}_\varphi  }$ is perfect  in the feature space, that is, it can map $\varphi ({\bf x})$ as $\varphi ({\bf x})$. The background information is not considered in this paper.

Experiments are conducted with both simulated sinusoidal signal  and captured DTV signal.  When the second order polynomial kernel with $c=1$ is used for kernel PCA approach, the experimental results show that kernel PCA is 4 dB better than PCA whether on the simulated signal or DTV signal. Kernel PCA can compete with EC method.  Kernel GLRT method is about 4 dB better than GLRT for DTV signal with appropriate choice of the width of Gaussian kernel's. Depending on the signal, kernel GLRT can even beat the EC method which owns  the perfect prior knowledge.

In this paper, the types of kernels and parameters in kernels are chosen manually by trial and error. How to choose an appropriate kernel function and parameter is still an open problem for us. 
In PCA and kernel PCA approaches, only the leading eigenvector is used for detection. Can both of the methods extend to the case that detection by subspaces consist of eigenvectors corresponding to nonzero eigenvalues? Motivated by kernel PCA approach, we know that a suitable choice of similarity measure is very important. What kind of similarity measure can be used for detection with the use of subspaces seems also an interesting and promising future direction.

\section*{Acknowledgment}
This work is funded by National Science Foundation
through two grants (ECCS-0901420 and ECCS-0821658), and
Office of Naval Research through two grants (N00010-10-1-
0810 and N00014-11-1-0006).




%


\bibliographystyle{ieeetr}
\bibliography{dsoref/spectrum_sensing,dsoref/cr_prediction}

\end{document}